\documentclass[
superscriptaddress,
preprint,
 amsmath,amssymb,
 aps,
]{revtex4-1} 
\usepackage{graphicx}
\usepackage{dcolumn}
\usepackage{setspace}

\begin{document}

\title{Dephasingless two-color terahertz generation}

\author{Tanner T. Simpson}
\email{tsim@lle.rochester.edu}
\affiliation{Laboratory for Laser Energetics, University of Rochester, Rochester, New York 14623-1299 USA}

\author{Jeremy J. Pigeon}
\affiliation{Laboratory for Laser Energetics, University of Rochester, Rochester, New York 14623-1299 USA}

\author{Kyle G. Miller}
\affiliation{Laboratory for Laser Energetics, University of Rochester, Rochester, New York 14623-1299 USA}

\author{Dillon Ramsey}
\affiliation{Laboratory for Laser Energetics, University of Rochester, Rochester, New York 14623-1299 USA}

\author{Dustin H. Froula}
\affiliation{Laboratory for Laser Energetics, University of Rochester, Rochester, New York 14623-1299 USA}

\author{John P. Palastro}
\email{jpal@lle.rochester.edu}
\affiliation{Laboratory for Laser Energetics, University of Rochester, Rochester, New York 14623-1299 USA}

\begin{abstract}
A laser pulse composed of a fundamental and an appropriately phased second harmonic can drive a time-dependent current of photoionized electrons that generates broadband THz radiation. Over the propagation distances relevant to many experiments, dispersion causes the relative phase between the harmonics to evolve. This ``dephasing'' slows the accumulation of THz energy and results in a multi-cycle THz pulse with significant angular dispersion. Here, we introduce a novel optical configuration that compensates the relative phase evolution, allowing for the formation of a half-cycle THz pulse with almost no angular dispersion. The configuration uses the spherical aberration of an axilens to map a prescribed radial phase variation in the near field to a desired longitudinal phase variation in the far field. Simulations that combine this configuration with an ultrashort flying focus demonstrate the formation of a half-cycle THz pulse with a controlled emission angle and 1/4 the angular divergence of the multi-cycle pulse created by a conventional optical configuration.
\end{abstract}

\flushbottom
\maketitle

\thispagestyle{empty}

\section*{Introduction}
The laser--matter interactions at the foundation of all nonlinear optical applications occur in dispersive media. These applications benefit from, and in some cases rely on, tailoring the wavefronts of a laser pulse to counteract dispersion. This is accomplished either through the structure of the medium or by preliminary structuring of the laser pulse. The former has become a mainstay of nonlinear optics: the different refractive indices along orthogonal crystal axes is regularly used for phase matching of sum-and-difference frequency generation \cite{midwinter1965effects, maker1962effects,boyd2019nonlinear}, while periodically poled lithium niobate allows for quasi-phase matching of the same processes \cite{fejer1992quasi, yamada1993first}. Structuring the laser pulse, although not as widespread, has found utility in optical rectification, where tilting the phase fronts with respect to the pulse front enables velocity matching to terahertz (THz) radiation \cite{hebling2002velocity, wu2023generation}. The recent development of more-advanced techniques for space--time structuring presents new opportunities to overcome the limitations imposed by dispersion and rethink the optimization of nonlinear optical applications \cite{kondakci2017diffraction,sainte2017controlling,froula2018spatiotemporal,kondakci2019optical,palastro2020dephasingless,caizergues2020phase,li2020optical,Jolly2020,Yessenov2022, liang2023space,ambat2023programmable,pigeon2024ultrabroadband,liberman2024use}.

A subset of space--time structuring techniques known as the ``flying focus'' modify the focal time and location of each frequency, temporal slice, or annulus of a pulse to produce an intensity peak that moves independently of dispersion\cite{froula2018spatiotemporal, palastro2020dephasingless,Jolly2020,simpson2020nonlinear,simpson2022spatiotemporal,li2024spatiotemporal}. With the ability to control the trajectory of an intensity peak, flying-focus techniques are predicted to enhance a wide range of applications and fundamental measurements, including advanced accelerators \cite{palastro2020dephasingless,ramsey2020vacuum,palastro2021laser, miller2023dephasingless}, radiation sources \cite{howard2019photon,franke2021optical,ramsey2022nonlinear, kabacinski2023spatio}, and signatures of quantum electrodynamical phenomena \cite{di2021unveiling,formanek2022radiation,formanek2024signatures}. Nevertheless, the flying-focus techniques that have been proposed \cite{simpson2020nonlinear,simpson2022spatiotemporal, li2024spatiotemporal} and experimentally demonstrated \cite{froula2018spatiotemporal,Jolly2020,pigeon2024ultrabroadband} thus far are limited to control over the \textit{intensity} of a laser pulse. Control over the \textit{phase}, which would substantially broaden the potential to enhance applications, has remained elusive and constrained by the medium. 

``Two-color'' THz generation in gases is a promising approach to producing high-quality, broadband THz radiation  \cite{kim2007terahertz,kim2008coherent,babushkin2010ultrafast,you2012off,berge20133d,johnson2013thz, zhang2018manipulation,yoo2019highly,koulouklidis2020observation,buldt2021gas,simpson2024spatiotemporal} that depends sensitively on both the intensity \textit{and} phase of a laser pulse. In this approach, a laser pulse composed of a fundamental and second harmonic ionizes a gas and drives a time-dependent current of freed electrons, which provides the source for the THz radiation. The strength of the current (and rate of THz generation) is proportional to the number of photoionized electrons and the drift velocity acquired at the instant of their birth, both of which are sensitive to the waveform (i.e., intensity and relative phase) of the two-color electric field \cite{kim2007terahertz}. Over short propagation distances, the relative phase between the harmonics is preserved, the strength and sign of the current are approximately constant, and the THz radiation constructively interferes at a Cherenkov angle determined by the THz phase velocity and ionization front velocity \cite{johnson2013thz,simpson2024spatiotemporal}. Over longer distances, dispersion causes the relative phase to evolve. This ``dephasing'' between the harmonics produces an oscillating current that disrupts the constructive interference, slows the rate of THz generation, and angularly disperses the THz bandwidth \cite{you2012off}.

Here we introduce a novel optical configuration that overcomes dephasing in two-color THz generation by simultaneously controlling the intensity \textit{and} relative phase of a two-color laser pulse. When the relative phase is precompensated to counteract dispersion, the two-color pulse drives a non-alternating current of photoionized electrons that produces a half-cycle THz pulse nearly devoid of angular dispersion. This ``dephasingless'' two-color THz generation is demonstrated by simulations that model the optical configuration, nonlinear propagation of the two-color pulse through gas, and the generation and propagation of the THz radiation. The optical configuration consists of three elements: an axilens to extend the focal range; a radially stepped echelon to control the focal time at each point within the focal range; and a diffractive lens, or ``phaser,'' to apply a weak radially varying phase to the second harmonic, which the axilens maps to a longitudinally varying phase (Fig. 1). The first two elements describe the ultrashort flying focus and provide control over the velocity of the intensity peak and emission angle of the THz \cite{simpson2024spatiotemporal}, while the novel phaser element provides the phase control. Aside from two-color THz generation, the intensity and phase control afforded by this configuration may benefit other nonlinear optical processes that rely on phase matching between multiple frequencies, such as harmonic generation and optical parametric amplification.

\begin{figure}
\centering
\includegraphics[scale = 1]{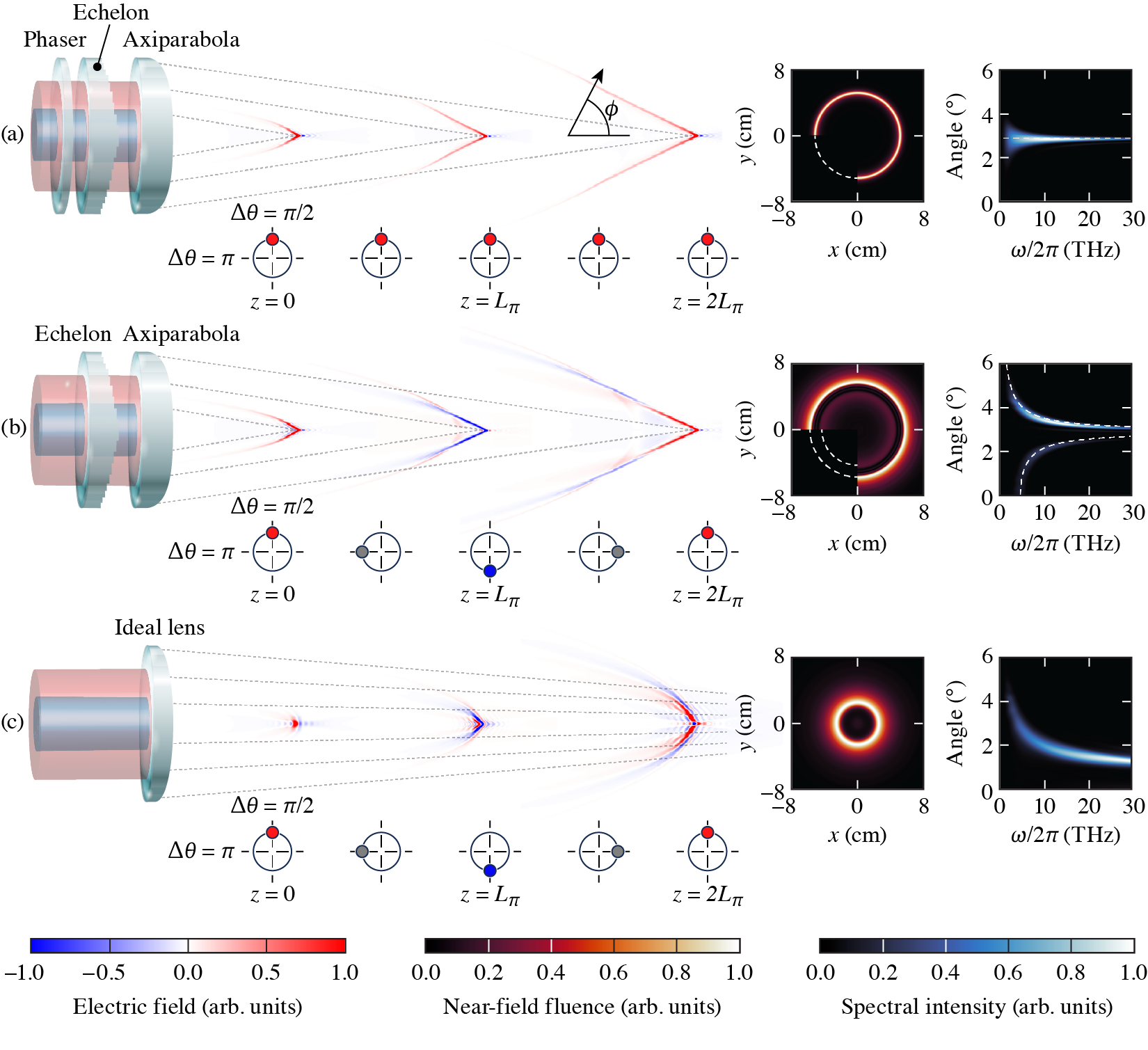}
\caption{Comparison of THz pulses generated by two-color laser pulses prepared with varying degrees of phase and intensity control. The spatiotemporal profiles for frequencies  $< 30$ THz are displayed to the right of each optical configuration at three longitudinal locations in the far field. The circles below each configuration illustrate the evolution of the relative phase between the harmonics of the two-color pulse $\Delta \theta$. (a) A phased-flying-focus pulse drives the conical emission of a half-cycle THz pulse over multiple dephasing lengths. The THz pulse is emitted into a single angle and has almost no angular dispersion. (b) A standard ultrashort-flying-focus pulse drives the conical emission of a multi-cycle THz pulse. Dephasing results in constructive interference at two characteristic angles and causes significant angular dispersion, which manifests in the near field as two diffuse rings. (c) A two-color laser pulse focused by an ideal lens also drives a multi-cycle THz pulse. In this case, the THz pulse evolves into a single diffuse ring in the near field, indicating both dephasing and complex evolution of the two-color laser pulse and resulting photocurrent.  The near-field transverse profiles (red circles) are shown 1 m beyond the edge of the gas, and the white dashed lines in (a) and (b) mark the predictions of Eqs. \eqref{eq:cherenkov} and \eqref{eq:cherenkovdephasing}. For ease of visualization, the optical configurations are displayed in transmission. In practice, only one harmonic is modified by the phaser.} 
\end{figure}

\section*{Results}
A laser pulse composed of a fundamental and second harmonic produces the largest current of photoionized electrons and maximizes the rate of THz generation when the relative phase between the harmonics is approximately $\pi/2$. While a relative phase of $\pi/2$ can occur locally, the frequency dependence of the refractive index due to the presence of gas and plasma causes the relative phase to evolve along the propagation path. During linear propagation, the relative phase advances continuously, reaching integer multiples of $\pi$ after every dephasing length
\begin{equation}
\label{eq:nominalL}
    L_\pi = \frac{\lambda_1}{4} \frac{1}{|n(\lambda_1)-n(\lambda_2)|},
\end{equation}
where $\lambda_1$ and $\lambda_2$ are the vacuum wavelengths of the two harmonics and $n$ is the linear refractive index of the gas. Structuring the phase and intensity of the two-color laser pulse to precompensate the evolution of the relative phase eliminates dephasing and ensures the formation of a half-cycle THz pulse devoid of angular dispersion (Fig. 1a).

To demonstrate the advantage of phase and intensity control in two-color THz generation, simulations were performed using the unidirectional pulse propagation equation\cite{kolesik2002unidirectional,kolesik2004nonlinear,couairon2011practitioner} initialized with pulses prepared by the optical configurations displayed in Fig. 1 (see Methods). In each simulation, a laser pulse composed of a fundamental and second harmonic was focused by the optical configuration in the near field to the entrance of the gas in the far field. The pulses propagated through and photoionized the gas, producing a current that generated THz radiation. The physical parameters for the simulations are provided in Table \ref{tab:table1}. The parameters chosen for the laser pulses and gas were motivated by commercially available high-repetition-rate Ti:Sapphire laser systems \cite{KMLabs} and typical two-color THz generation experiments \cite{you2012off, kim2008coherent, yoo2019highly, koulouklidis2020observation}. 

Figure 1 compares the THz pulses resulting from three optical configurations, each of which provides a different degree of control over the waveform of the two-color laser pulse. The novel, phased-flying-focus design introduced in this work (Fig. 1a) precompensates the relative phase of the two harmonics to counteract dispersion and produces an intensity peak that travels at a specified velocity $v_f$. By counteracting dispersion, the two-color pulse drives the conical emission of a half-cycle THz pulse over multiple dephasing lengths. The pulse is emitted into an angle determined by the Cherenkov condition
\begin{equation}
\label{eq:cherenkov}
    \phi = \arccos(v_T/v_f), 
\end{equation}
where $v_T$ is the phase velocity of the THz radiation. The emission angle $\phi$ has only a weak dependence on frequency through the phase velocity $v_T$. As a result, all of the generated THz frequencies are emitted into nearly the same angle, which can be controlled through the focal velocity $v_f$. The minimal angular dispersion results in a near-field fluence profile characterized by a single, narrow ring.

The standard flying focus and ideal lens configurations do not precompensate the relative phase of the two harmonics (Figs. 1b and c). The standard flying focus does, however, offer control over the velocity of the intensity peak. In this case (Fig. 1b), the two-color pulse drives the conical emission of a multi-cycle THz pulse characterized by two emission angles \cite{you2012off,johnson2013thz} 
\begin{equation}
\label{eq:cherenkovdephasing}
    \phi_{\pm} = \arccos\left( \frac{v_T}{v_f} \pm \frac{\pi v_T}{\omega_TL_\pi} \right), 
\end{equation}
where $\omega_T$ is the THz frequency. Compared to $\phi$ [Eq. \eqref{eq:cherenkov}], the emission angles $\phi_{\pm}$ have a stronger dependence on the THz frequency---the second term on the right-hand side of Eq. \eqref{eq:cherenkovdephasing}. This additional frequency dependence is a direct consequence of dephasing and causes significant angular dispersion. The resulting near-field profile exhibits two rings corresponding to $\phi_{\pm}$, each of which are broadened by the angular dispersion. With neither phase nor intensity control, a two-color laser pulse focused by the ideal lens drives a multi-cycle THz pulse with a complex and diffuse angular emission pattern that is not accurately predicted by Eqs. \eqref{eq:cherenkov} or \eqref{eq:cherenkovdephasing} (Fig. 1c). The diffuse near-field emission pattern indicates both dephasing and intensity evolution due to the complex interplay between linear focusing, self-focusing, and plasma refraction. 

\begin{table}[t]
\centering
\caption{\label{tab:table1}%
Simulation parameters for the two-color laser pulse, optical configurations, and gas. 
}
\begin{tabular}{lc}
\hline
\textrm{Laser Parameters}&
\textrm{Value}\\
\hline
$\lambda_1 (\mu \mathrm{m}$) & $0.8$ \\
$\lambda_1$ Power (GW) & 100\\
$\lambda_1$ Duration (fs) & 30\\
$\lambda_2 (\mu \mathrm{m})$ & $0.4$ \\
$\lambda_2$ Power (GW) & 20\\
$\lambda_2$ Duration (fs) & $15\sqrt{2}$\\
Total energy (mJ) & 3.5\\
\hline
\textrm{Optics Parameters}&
\textrm{Value}\\
\hline
Axiparabola $f/\#$ & 15\\
Axiparabola radius $R$ (cm) & 5 \\
Focal range $L$ (cm) & 12\\
Phaser length $L_P$ (cm) & 2.5\\
Focal velocity $v_f$ (c) & 1.001 \\
\hline
Ideal lens $f/\#$ & 250\\
Ideal lens confocal parameter (cm) & 12\\
\hline
\textrm{Medium Parameters}&
\textrm{Value}\\
\hline
Species & Ar\\
Density ($\mathrm{cm}^{-3}$) & $2.7{\times}10^{19}$\\
Nominal dephasing length $L_\pi$ (cm) & 2.9\\
Electron-neutral collision frequency (THz) & 10\\
Nonlinear refractive index ($ \mathrm{cm}^2/\mathrm{W}$) \cite{zahedpour2015measurement} & $1{\times}10^{-19}$ 
\end{tabular}
\end{table}

The optical configuration for the phased flying focus consists of three elements: an axiparabola, a radially stepped echelon, and a weak diffractive lens or ``phaser.'' The axiparabola, a specific type of axilens, is designed to produce a uniform, on-axis intensity over an extended distance when illuminated by a laser pulse with a flattop transverse profile \cite{sochacki1992nonparaxial,smartsev2019axiparabola, palastro2020dephasingless,oubrerie2022axiparabola, ambat2023programmable}. Light rays incident at different radii $r$ on the axiparabola are focused to different longitudinal locations $z = f(r)$, where
\begin{equation}
\label{eq:axiparabola}
f(r) = f_0+\left(\frac{r}{R}\right)^2L,
\end{equation}
$f_0$ is the nominal focal length, $L$ is the focal range, and $R$ is the maximum radius of the optic. The radially stepped echelon applies a radially dependent delay $\tau(r)$ that controls the arrival time of the light rays at their focus \cite{palastro2020dephasingless, ambat2023programmable}. For a desired velocity $v_f$,
\begin{equation}
\label{eq:echelon}
\tau(r) = \frac{R^2}{2cL}\left[ \frac{f(r)}{f_0} - 1 + \ln\Bigl(\frac{f(r)}{f_0}\Bigr) \right] + \frac{L}{cR^2}\left(1-\frac{v_f}{c}\right)r^2.
\end{equation}
To avoid distorting the phase front applied by the axiparabola, the echelon applies this delay in discrete concentric rings with varying width and half-integer-wavelength depth. Together, the axiparabola and echelon produce an ultrashort moving focus, which travels at a velocity $v_f$ that is independent of dispersion \cite{palastro2020dephasingless,ambat2023programmable,  pigeon2024ultrabroadband}.

The weak diffractive lens, or phaser, imparts the necessary phase to precompensate the relative phase advance of the first and second harmonic due to dispersion in the gas. To design such an optic, recall that the axiparabola maps each radius in the near field $r$ to a different on-axis location in the far field $z=f(r)$. As a result, a desired longitudinal phase variation in the far field $\psi(z)$ can be mapped back to a required radial phase variation $\psi[z(r)]$ in the near field. A longitudinal phase variation of 
\begin{equation}\label{eq:phaserz}
    \psi_P(z) = \pi \frac{z}{L_{\pi}},
\end{equation}
applied to one of the harmonics would result in their relative phase advancing by $\pi$ after every dephasing length, which is exactly the advance required to cancel the phase evolution due to dispersion. Using $f(r)$, Eq. \eqref{eq:phaserz} can be reexpressed as the near-field radial phase variation
\begin{equation}\label{eq:phaserr}
    \psi_P[z(r)] = \pi\left(\frac{r}{R}\right)^2 \frac{L}{L_{\pi}},
\end{equation}
where constant phase terms have been dropped. This phase is equivalent to that applied by a diffractive lens with a focal length $f_P = R^2L_{\pi}/\lambda_2 L$. In practice, the actual dephasing length depends on the detailed propagation and includes contributions from the nonlinear refractive index of the gas and plasma formation. Thus, in order to optimize the phaser optic for THz generation, it is useful to replace the nominal dephasing length $L_{\pi}$ in Eq. \eqref{eq:phaserr} or $f_P$ with an adjustable length $L_P$. For the parameters considered here, the formation of a half-cycle THz pulse was not sensitive to the precise value of $L_P$ and setting $L_P = L_{\pi}$ yielded similar, albeit slightly less optimized, results (see Methods).

\begin{figure}[t]
\centering
\includegraphics[scale = 1]{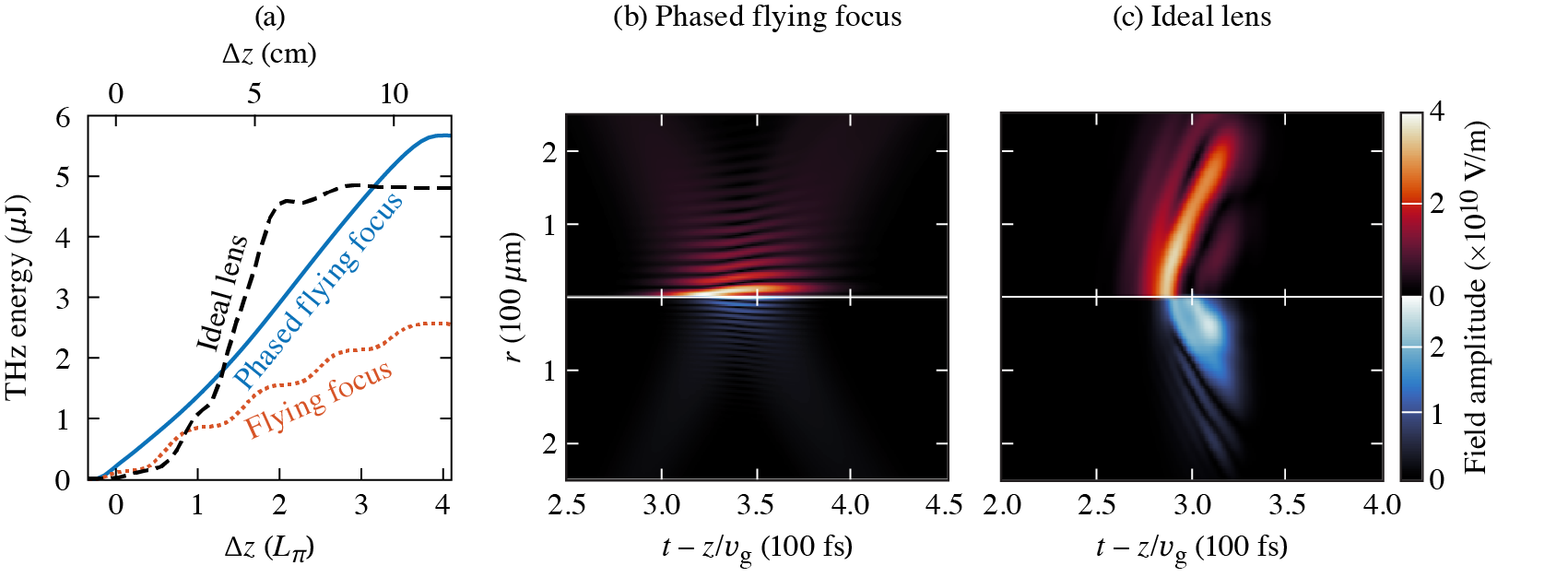}
\caption{The accumulation of THz energy ($<30$THz) with distance depends on the degree of phase and intensity control and nonlinear propagation of the two-color pulse. (a) THz energy as a function of distance for the phased flying focus (blue, solid), standard ultrashort flying focus (red, dotted), and a conventional pulse focused by an ideal lens (black, dashed). (b,c) The electric field amplitudes of the first (top) and second harmonic (bottom) for (b) the phased flying focus and (c) conventional pulse at $\Delta z = 6 \; \mathrm{cm}$. The flying-focus pulses are relatively resistant to nonlinear propagation, which keeps the first and second harmonic overlapped in space and time. This allows the THz energy to grows continuously when dephasing is eliminated with the phaser element (blue, solid) or periodically when dephasing is present (red, dotted). The first harmonic of the conventional pulse undergoes rapid self-focusing, which causes a loss of spatiotemporal overlap between the first and second harmonic. As a result, the THz energy grows over a short distance and saturates. }
\end{figure}\label{fig:f2}

Figure 2a displays the energy of the THz pulses produced by the three optical configurations as a function of distance. When both the relative phase and intensity of the two-color pulse are controlled using the phased flying focus (Fig. 1a), the THz energy increases at a near-constant rate over $\sim$four dephasing lengths. This steady increase in the THz energy indicates the absence of dephasing. When only the intensity is controlled using the ultrashort flying focus without the phaser element (Fig. 1b), the energy grows where the relative phase between the two harmonics is favorable for driving a net current and plateaus where the relative phase is unfavorable. This energy growth is consistent with a simple theoretical picture of dephasing in two-color THz generation \cite{you2012off}. When neither phase nor intensity control are applied, the THz energy grows rapidly over a short distance and then saturates (Fig. 1c). 

The rapid growth and saturation of the THz energy observed with the conventional pulse is a result of nonlinear propagation. In each configuration, the power of the two-color pulse exceeds the critical power for self-focusing: $P/P_c = 10$ and $8$ for the fundamental and second harmonic. Upon entering the gas, the fundamental of the conventional pulse rapidly self-focuses, which increases its intensity. Once the earlier time slices of the fundamental reach sufficient intensity to ionize the gas, they create a plasma, which refracts subsequent time slices. This refraction hollows out the back of the fundamental (Fig. 2c). The second harmonic, in contrast, is less susceptible to plasma refraction and approximately maintains its spatiotemporal profile. The loss of overlap between the two harmonics within the ionization front terminates the THz generation.

The flying-focus pulses are relatively resistant to nonlinear propagation. This can be understood by considering two properties of ultrashort-flying-focus pulses. First, the transverse profile of these pulses features a central maximum surrounded by multiple rings. Thus, even though the total power is high, the power contained within the central maximum is low, which mitigates self-focusing \cite{miller2023dephasingless} (a similar property leads to larger effective critical powers for higher-order Laguerre Gaussian beams and partially incoherent beams \cite{Sa2019,Bang1999}). Second, the $f/\#$ of the flying-focus configurations is much smaller, so that each annulus of the pulse comes in and out of focus relatively quickly and does not have sufficient propagation path at high intensity to accumulate a significant time-dependent, nonlinear phase.

\begin{figure}[t]
\centering
\includegraphics[scale = 1]{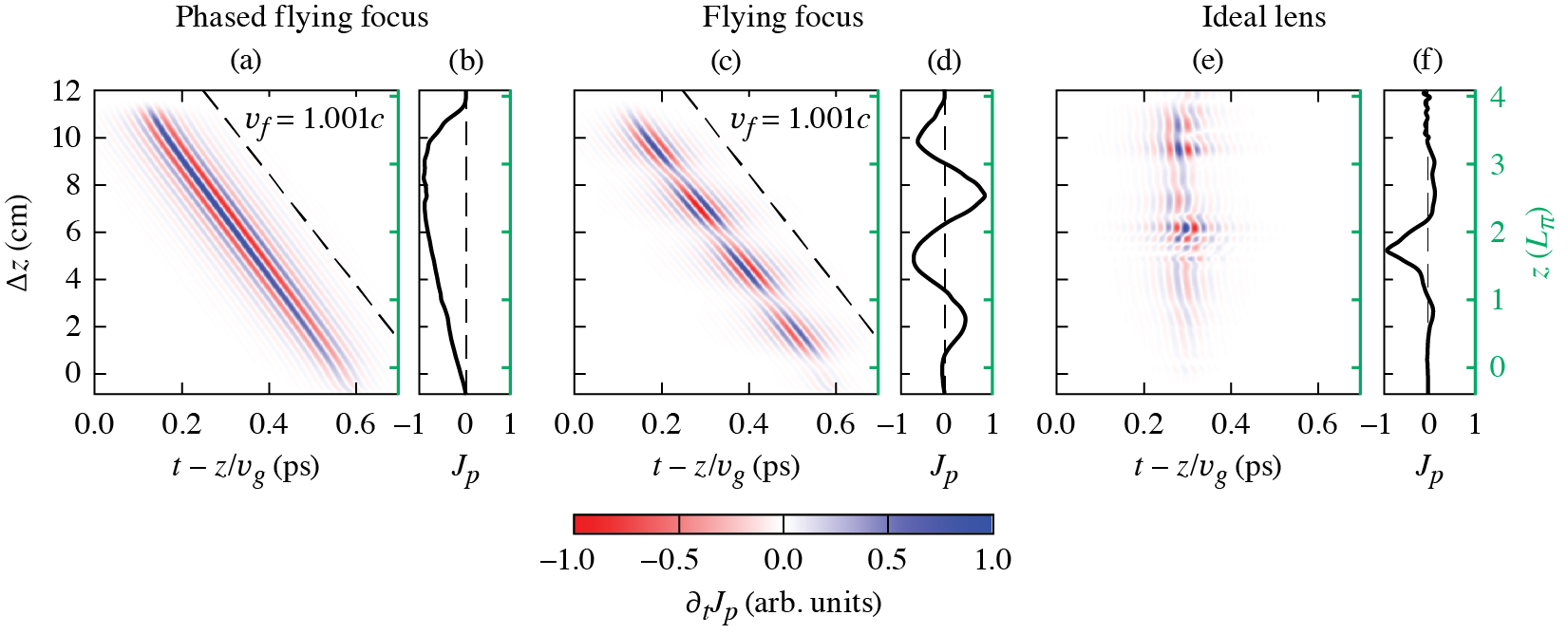}
\caption{Evolution of the differential current densities $\partial_t J_p$ and current densities $J_p$ produced by the (a,b) phased flying focus, (c,d) standard ultrashort flying focus, and (e,f) ideal lens configurations. The phased flying focus drives a differential current with a near-constant temporal profile that advances at the prescribed focal velocity $v_f = 1.001$ over $\sim$four dephasing lengths. The near-constant temporal profile of the differential current yields a current that does not oscillate with distance. The differential current densities have been low-pass filtered at 30 THz, and the current density $J_p$ is shown $500$ fs after the maximum of $\partial_t J_p$. Note that in the moving frame $t-z/v_g$, the focus follows the trajectory $\Delta z = v_f(1-v_f/v_g)^{-1}(t-z/v_g)$. }
\end{figure}\label{fig:f3}

Figure 3 illustrates the effects of propagation and dephasing on the photocurrent responsible for THz generation. The differential current density $\partial_t J_p$ is the direct source of the THz field (Figs. 3a, 3c, and 3e), while the net current density $J_p$ behind the two-color pulse indicates whether or not THz was generated (Figs. 3b, 3d, and 3f). The differential current density depends on the drift velocity acquired by ionized electrons at the time of their birth $v_d$ and the ionization rate: $\partial_t J_p \sim ev_d \partial_t n_e$, where $n_e$ is the electron density. The duration over which the two-color pulse ionizes the gas determines the temporal envelope, or width $\tau$, of $\partial_t J_p$. The two-color pulse produces a net current with a bandwidth $\sim$$1/\tau$ sufficient to drive THz radiation when the integral of $\partial_t J_p$ over the duration of the pulse (i.e., $J_p$) is nonzero.

The evolution of $\partial_t J_p$ and $J_p$, both in time and along the propagation path $\Delta z$, determine the structure of the THz pulse. The differential current driven by the phased flying focus has a near-constant temporal profile that advances at the prescribed focal velocity $v_f = 1.001c$ over the entire interaction length $L\sim4$$L_{\pi}$ (Fig. 3a). The constant amplitude of $\partial_t J_p$ along lines $\Delta z = v_f t + z_0$ demonstrates the absence of dephasing. The near-constant profile and absence of dephasing yield a non-alternating current $J_p$ (Fig. 3b) that facilitates the constructive interference of the THz radiation into a half-cycle burst emitted at the Cherenkov angle [Eq. \eqref{eq:cherenkov} and Fig. 1a].

The differential current driven by the standard flying focus evolves as it advances at the focal velocity (Fig. 3c). In this case, the amplitude oscillates along lines $\Delta z = v_f t + z_0$ at a spatial period approximately equal to the nominal dephasing length $L_{\pi}$. This yields an alternating current $J_p$ (Fig. 3d) that generates a multi-cycle THz pulse characterized by two angles and significant angular dispersion [Eq. \eqref{eq:cherenkovdephasing} and Fig. 1b]. The differential current driven by the conventional pulse also evolves across the interaction region (Fig. 3e). Here, the amplitude has a complex temporal dependence along lines $\Delta z = v_g t + z_0$ due to dephasing and nonlinear propagation of the two-color pulse. Changes in the local velocity of the current density result in the emission of a multi-cycle THz pulse into range of angles, each of which are broadened by angular dispersion (Fig. 1c). 

For both the phased and standard flying focus, the amplitude of the current density $J_p$ grows with distance within the focal range (Figs. 3b and 3d). This is primarily due to the spherical aberration of the axiparabola, which causes the incidence angle to increase with distance within the focal range. Rays incident on the plasma at a steeper angle penetrate further into the plasma before defocusing. As a result, the pulse can reach a higher intensity, ionize more electrons, and produce a larger photocurrent.
 
\section*{Discussion}
While the THz pulses resulting from the phased flying focus and ideal lens configurations had comparable energy (Fig. 2), further optimization of the phased flying focus is possible. The energy of the THz pulse produced by the phased flying focus can be increased by either (i) extending the focal range or (ii) using an axiparabola with a smaller $f/\#$. Extending the focal range requires increasing the energy of the two-color pulse to maintain a fixed peak intensity and mitigating the loss of spatiotemporal overlap between the two harmonics due to group velocity dispersion (GVD). For the parameters simulated here, GVD causes the first and second harmonic to walk off from one another at a rate of $\approx$70 fs/m or $\approx$8.4 fs over $L = 12\;\mathrm{cm}$, which is already comparable to the duration of the second harmonic $\tau_2 = 21 \; \mathrm{fs}$. This walkoff can be mitigated by increasing the duration of the first harmonic. The increased duration would ensure that the harmonics remain overlapped over a longer distance without affecting the rise time of the ionization front responsible for THz generation, which is set by the duration of the second harmonic.

The axiparabola used for the phased and standard flying focus had a full-aperture $f/\# = 15$. Simulations (not shown) were also conducted for an $f/\# = 20$ axiparabola with all other parameters held constant. Qualitatively, the results were nearly identical: The phased flying focus produced a single-cycle THz pulse with almost no angular dispersion, and the standard flying focus produced a multi-cycle pulse with a near-field profile that exhibited two diffuse rings. Quantitatively, the $f/\# = 15$ configuration resulted in ${\sim}2\times$ greater THz energy than the $f/\# = 20$ configuration. This is due to two effects. First, as shown in Ref. \cite{ambat2023programmable}, smaller $f/\#$'s mitigate the elongation of the pulse duration and accompanying drop in intensity caused by the chromatic aberration imparted by the echelon. Second, a smaller $f/\#$ increases the incidence angle of the two-color pulse, which allows the pulse to penetrate further into the plasma and reach a higher intensity. The shorter duration results in more broadband THz generation, while the higher peak intensity increases the rate of ionization, thereby enhancing the THz-generating current. As a result, use of an $f/\# < 15$ is expected to further enhance the THz energy. 

The weak diffractive lens or phaser can compensate the relative phase evolution of the first and second harmonic over many dephasing lengths, but there is a limit: if the applied phase becomes too large, it can modify the focal velocity and lengthen the duration of the affected harmonic. In the case of the second harmonic, the phaser extends the focal range by a distance $L_D = (\Delta \lambda / \lambda_2)f_0^2/f_P$ and imparts a nonideal radial delay $\tau_D(r) = -r^2/2cf_P$, where $\Delta \lambda$ is the bandwidth of the second harmonic and $f_0 \ll f_P$ has been assumed. The phaser will have a negligible effect on the velocity of the focus and temporal duration $\tau_2$ when $L_D \ll L$ and $|\tau_D| \ll |\tau|$ [see Eqs. \eqref{eq:axiparabola} and \eqref{eq:echelon}]. For the parameters in Table \ref{tab:table1} where $L \sim4$ dephasing lengths, these conditions are readily satisfied: $L_D=2 \times 10^{-3}\;\mathrm{cm}$ and $|\tau_D(R)| = 3\;\mathrm{fs}$, while  $L = 12\; \mathrm{cm}$ and $|\tau(R)| = 5\; \mathrm{ps}$. 

A novel optical configuration that enables simultaneous control over the intensity and relative phase of a two-color laser pulse can eliminate dephasing in two-color THz generation. Simulations that modelled this configuration, the nonlinear propagation of the two-color laser pulse through gas, and the generation and propagation of THz radiation demonstrated the formation of a single-cycle THz pulse with a controlled emission angle and almost no angular dispersion. This contrasts the multi-cycle, angularly dispersed THz pulse that formed when the two-color pulse was focused by an ideal lens. The novel configuration combines the axiparabola and echelon of an ultrashort flying focus \cite{palastro2020dephasingless,ambat2023programmable,pigeon2024ultrabroadband} with a weak diffractive lens (or equivalent chromatic assembly). The axiparabola and echelon provide control over the motion of the far-field intensity, while the axiparabola and diffractive lens provide control over the relative phase evolution. The near absence of angular dispersion and controlled emission angle of the THz pulses produced by this phased flying focus can facilitate collection and refocusing of the radiation for subsequent applications. Future work will explore whether the intensity and phase control afforded by this configuration can be applied to other nonlinear optical processes that rely on the interaction of multiple frequencies and whether coatings or a redesign of the echelon can be used to impart an equivalent phase difference $\psi_P$ between the two harmonics. This would eliminate the need for an additional optic and allow the first and second harmonic to travel a common path.

\section*{Methods}
The optical configuration, pulse propagation, and two-color THz generation were simulated as follows. The transverse electric field of the laser pulse was initialized in the near field with two carrier frequencies---a first and second harmonic, each with the same transverse profile. The analytical phases of the optical configurations illustrated in Fig. 1 were then applied to the initial electric fields of each harmonic in the frequency domain. With the phases applied, the two-color laser pulses were propagated in vacuum from the exit of the optical configuration to the beginning of the gas using the frequency-domain Fresnel integral.\cite{palastro2018ionization,simpson2024spatiotemporal} The electric fields resulting from the Fresnel integral provided the initial condition for the unidirectional pulse propagation equation (UPPE).\cite{kolesik2002unidirectional,kolesik2004nonlinear,couairon2011practitioner} The UPPE was used to simulate the nonlinear propagation of the two-color laser pulse through the gas and the generation and propagation of the THz radiation.

For the phased-flying-focus configuration, the analytical phase applied in the near field included terms for the phaser, echelon, and axiparabola: $\psi = \psi_{P} + \psi_{E} + \psi_{AP}$. The contribution from the phaser $\psi_P$ is given by Eq. \eqref{eq:phaserr} with $L_\pi = L_P$ for the second harmonic and $\psi_P=0$ for the first harmonic. The contributions from the echelon and axiparabola are given by 
\begin{equation}
    \psi_E = -\frac{\omega}{c}\frac{\lambda_1}{2}\left(\mathrm{ceil}\left[\frac{c\tau(r)}{\lambda_1}\right]+\mathrm{floor}\left[\frac{c\tau(r)}{\lambda_1}\right]\right),
\end{equation}
where $\tau(r)$ is found in Eq. \eqref{eq:echelon}, and
\begin{equation}
    \psi_{AP} = -\frac{\omega}{c}\frac{R^2}{2L}\ln\left(1+\frac{Lr^2}{f_0R^2}\right).
\end{equation}
A single echelon can be and was used for both harmonics because their wavelengths are integer multiples $\lambda_1=2\lambda_2$. The standard flying-focus configuration used the same $\psi_E$ and $\psi_{AP}$, but did not include the phaser element (i.e., $\psi_P=0$ for both harmonics). For the ideal lens configuration, the analytical phase included a single term for the lens: $\psi = \psi_l$, where
\begin{equation}
    \psi_l = -\frac{\omega}{c}\frac{r^2}{2f_l}
\end{equation}
and $f_l$ is the focal length.

The parameters describing the optical configurations and the initial field profiles are provided in Table \ref{tab:table1}. For the ideal lens configuration, the confocal parameter of the lens was set equal to the focal range of the flying focus to ensure a comparable length of high-intensity propagation and THz generation. The ideal lens $f/\# = 250$ was acheived using a $f_l=5 \;\mathrm{m}$ focal length and initial spot size $w_i = 1\; \mathrm{cm}$. The transverse profiles of the pulses were Gaussian and flattop for the ideal lens and flying-focus configurations, respectively. The laser parameters were chosen based on commercially available, high-repetition-rate Ti:sapphire laser systems. Second harmonic generation in crystals like beta barium borate results in a second harmonic pulse with a duration that is $\sim$$\sqrt{2}/2$ shorter than the first harmonic. This was accounted for when setting the initial duration of the second harmonic.

In the figures, $\Delta z=0$ corresponds to the nominal focal point of the axiparabola $z = f_0$ for the flying focus pulses and $z = f_l - 4.5L_\pi$ for the conventional pulse. The shift of $4.5L_\pi$ optimized the THz generation for the conventional pulse. The initial relative phase and phaser length $L_P$ were tuned over multiple simulations to produce the maximum THz energy. This occurred for 
$L_P=2.5\,\mathrm{cm}$, which matched the observed dephasing length. However, setting the phaser length equal to the nominal dephasing length determined solely by the dispersion of the background gas (i.e., $L_P = L_\pi = 2.9\,\mathrm{cm}$) was sufficient to significantly mitigate dephasing and produce a near-half cycle THz pulse. In a series of simulations with lower energy but otherwise equivalent two-color pulses (not shown), the observed dephasing length approached the nominal dephasing length $L_\pi$ as the pulse energy was decreased: the lower-energy pulses ionized fewer electrons, which reduced the plasma contribution to the dispersion. In an experiment, the gas pressure could be adjusted to match a specific phaser design or to compensate other uncertainties in experimental conditions.

The Fresnel integral was used to propagate from the optical configuration  to the start of the interaction region because it can have a lower computational cost than the UPPE when the $f /\#$ is small. More specifically, the Fresnel integral can take advantage of different radial grids in the near field (optics plane) and far field (start of the interaction region) without a significant increase in implementation complexity. The UPPE captures linear dispersion to all orders and can simulate arbitrarily large frequency ranges, making it ideal for modeling nonlinear propagation of electromagnetic waves with diverse frequency content. As implemented here, the UPPE included source terms for the third-order nonlinear polarization density due to bound electrons, an effective current density that accounts for ionization energy losses, and the current density of the photoionized electrons with damping from electron-neutral collisions (see Ref. \cite{simpson2024spatiotemporal}). The linear dispersion was calculated using the Sellmeier equation found in Ref. \cite{peck1964dispersion} and extrapolated to THz frequencies. 

The UPPE was solved in a cylindrically symmetric, 2D+$t$ geometry with a second-order predictor-corrector method for the nonlinear source terms. The temporal resolution and domain sizes were $\Delta t=70 \,\mathrm{as}$ and $T = 2.4\,\mathrm{ps}$, corresponding to a spectral resolution of $\Delta f= 0.5 \, \mathrm{THz}$. The minimum radial resolution was $\Delta r = 3\,\mu\mathrm{m}$ with a radial domain size $r_{\mathrm{max}} = 1\,\mathrm{cm}$. The axial step size was $\Delta z = 20 \;\mu\mathrm{m}$. The resolution was adjusted to ensure convergence, and the maximum bounds on $t$ and $r$ were chosen to mitigate aliasing and spurious reflections, respectively. 

\bibliography{biblio}

\section*{Acknowledgements}

The authors would like to thank M. V. Ambat, A. Elliot, D. Gitlin, L. Mack, H. Markland, and M. L. P. Chong for productive discussions. 

This report was prepared as an account of work sponsored by an agency of the U.S. Government. Neither the U.S. Government nor any agency thereof, nor any of their employees, makes any warranty, express or implied, or assumes any legal liability or responsibility for the accuracy, completeness, or usefulness of any information, apparatus, product, or process disclosed, or represents that its use would not infringe privately owned rights. Reference herein to any specific commercial product, process, or service by trade name, trademark, manufacturer, or otherwise does not necessarily constitute or imply its endorsement, recommendation, or favoring by the U.S. Government or any agency thereof. The views and opinions of authors expressed herein do not necessarily state or reflect those of the U.S. Government or any agency thereof.

This material is based upon work supported by the Office of Fusion Energy Sciences under Award Numbers DE-SC0021057, the Department of Energy National Nuclear Security Administration under Award Number DE-NA0004144, the University of Rochester, and the New York State Energy Research and Development Authority.

\section*{Author contributions statement}
T.T.S., D.H.F., and J.P.P. conceived the idea of phase control for coherent THz generation. T.T.S. performed and analyzed the simulations. T.T.S., K.G.M., D.R., and J.P.P. developed the theoretical and simulation techniques. J.J.P. and D.H.F. provided experimental insights. T.T.S. and J.P.P. wrote the manuscript. All authors reviewed and edited the manuscript. 

\section*{Additional information}
The authors declare no competing interests. 

\end{document}